\documentclass{sig-alternate}
\usepackage[latin1]{inputenc}
\usepackage{url}

\begin{document}

\title{Browser-based distributed evolutionary computation: performance and scaling behavior}
%\thanks{Supported by project TIC2003-09481-C04.}

\numberofauthors{3}
\author{\alignauthor Juan-J. Merelo\\
	Antonio Mora-Garcia\\
	JLJ Laredo\\
       \affaddr{Dpto. Arquitectura y Tecnologia de Computadores}\\
       \affaddr{ETS Ingenieria Informatica}\\
       \affaddr{Universidad de Granada, Granada 18071}\\
       \email{\{jmerelo|amorag|juanlu\}@geneura.ugr.es}
\alignauthor Juan Lupion\\
       \affaddr{The Cocktail}\\
       \email{pantulis@gmail.com}
\alignauthor Fernando Tricas\\
       \affaddr{Dpto. Informatica e Ingenieria de Sistemas}\\
       \affaddr{Centro Politecnico Superior}\\
       \affaddr{Universidad de Zaragoza}\\
       \email{ftricas@unizar.es}
}

%\numberofauthors{1}
%\author{
%\alignauthor John Doe\\
%       \affaddr{Miskatonic University}\\
%       \email{ctulhy@miskatonic.edu}}

\date{\today}
\maketitle
\begin{abstract}
The challenge of ad-hoc computing is to find the way of taking
advantage of spare cycles in an efficient way that takes into account
all capabilities of the devices and interconnections available to
them. In this paper we explore distributed evolutionary computation based on the
Ruby on Rails framework, which overlays a Model-View-Controller on
evolutionary computation. It allows anybody with a web browser (that is,
mostly everybody connected to the Internet) to participate in an
evolutionary computation experiment. Using a straightforward farming model,
we consider different factors, such as the size of the population
used. We are mostly interested in how they impact on performance, but also
the scaling behavior when a non-trivial number of computers is applied to
the problem. Experiments show the impact of different packet sizes on
performance, as well as a quite limited scaling behavior, due to the
characteristics of the server. Several solutions for that problem are
proposed. 
\end{abstract}


%*******************************************************
%*******************************************************
\section{Introduction}
%*******************************************************
\label{sec:intro}

Application--level networks, ALNs, are configured as a set of clients that
can provide their spare CPU cycles by means of an application that  can be
downloaded, establishing a distributed computation network. Some ALN like SETI@Home have been quite successful, while other
experiments such as Popular Power have not. 
Many of these ALNs provide spare or ad hoc computational power for
distributed computing experiments.

The key feature of these application--level networks is the simplicity of
use: we believe that the best way to obtain the participation   of as many users as possible is to avoid trouble. In particular, it will be easier if they do not need to
download a special application to participate.
For this reason, we are exploring the use of elements that are usually
installed in the user's computer; in this sense, it is clear that the
web browser is an element almost universally installed: it is available
even in some cellular phones.
Moreover, most browsers  include a JavaScript
interpreter~\cite{Gilorien:2000:DJ,Shah:1996:BGJ,js:reference}. JavaScript is an interpreted
language, initially proposed  by Netscape, and later adopted as an ECMA
standard \cite{ECMA-262,ECMA-290,ECMA-327,ECMA-357}. 
In this way, most browsers are compatible, at least at a language level
(not always at the level of browser objects, where there exists a
reasonable compatibility, anyway).

The ability to use these features for distributed computing appeared with
the {\sf XmlHttpRequest} object, which allows asynchronous petitions to the
server, in what has been called AJAX, Asynchronous JavaScript and
XML~\cite{wiki:AJAX:en}\footnote{AJAX is just one of the possible ways to
perform asynchronous client-server communication, the others being
AJAJ (Asynchronous Javascript and JSON), and {\em remoting} using
applets or embedded objects. However, it is quite popular, and a wide
user base and documentation is available for it.}. The traditional client/server model becomes then
more egalitarian, or closer to a peer to peer model, since a
bidirectional communication line appears: the browser can make calls to the
server, do some computation and later send the results to the server.
The proposed mechanism is as follows: the {\sf XmlHttpRequest} is provided
with a request to the server and a pointer to a {\em callback} function. 
The request generates an event, which is asynchronously activated when a
reply is received  making use of the
 {\em callback} function. 
Following this approach the browser is not locked, providing the way to
program applications that are similar to  the ones the volunteers are used
to, in the sense that they do not have to wait for the application to load
and render the whole screen every time a request is made.
On the other side, this provides a way to use the browser for application
level networks and its use for distributed computing systems,
since the request-response loop does not need the user participation in a
fashion very similar  to any other distributed computing application.
This feature can be controlled from the server with any programming
language. Of course, it can also be combined with other distributed
programming frameworks based on OpenGrid~\cite{ogsa}. 

The server can be programmed traditionally using any of the paradigms
available (servlets or CGIs, for instance), but in order to produce a
rapid development of the application, the use of Ruby 
on Rails \cite{programming:ruby,RoR:assessing,RoR:Agile} was
considered ~\cite{dconrails-jp}. 
It is a framework based on Ruby language and in the 
Model/View/Controller~\cite{gamma93design,MVC-Cookbook} paradigm
(which has been used
before in evolutive computing; for example, in~\cite{mvc-ec}).
In this context the data model is clearly separated (usually with a
database management system) from the different {\em views} 
(HTML templates that will be used by the server to fill in the data and
send the pages to the client), 
and from the control part, the functions that modify and manage data; 
in RoR, the controllers are in charge of receiving client requests and
react to them.
When constructing a RoR application  we need to setup a data model, a
controller which will distribute the computation among client an server and
a set of views. 
In our case, these views will include the client's computation part (since
the JavaScript programs are included in the pages that will be served to
clients)\footnote{Please note that this is only one possible
arrangement, which was deemed the simplest for these initial
studies. Others, including storing JS code in the database (making it
part of the model) are also possible within RoR}.

We are concentrating on distributing, evolutionary computation
applications, which has already  been adapted to several paradigms of
parallel and distributed computing
 (for example, Jini \cite{Jini:Paralelos}, JavaSpaces
\cite{Setzkorn:2004:KELSI}, Java with applets \cite{chong99}, MPI \cite{PEO,javi:jp2001},
service oriented architectures \cite{jj:jp2001,876712} and P2P \cite{maribel:jp2001,viktor:cedi2005})
and it is adequate for this kind of exercise for several reasons:
it is a population based method, so computation can be distributed
among nodes in
many different ways; besides,
some works suggest that there are synergies among evolutive algorithms and
parallelization: isolated populations that are connected only eventually
avoid the lost of diversity and produce better solutions in fewer time
obtaining, in some cases, superlinear accelerations~\cite{Alba99Pue}. 

This work is more than a proof of concept~\cite{dconrails-jp}: first,
we will try to establish a baseline for the performance of a
JavaScript-based evolutionary algorithm by running benchmarks on
several virtual machines; then, we will try to see how different
elements of the system, especially latency, influence performance,
and, finally, we will do some measurements of DCoR system
({\em Distributed Computation on Rails}) in real network, in order
to see how this computation scales.

The rest of the paper is organized as follows: next, an exposition of
the state of the art in volunteer and so-called {\em parasitic}
computing is presented. Section \ref{sec:met} presents the DCoR
(Distributed Computation on Rails) system; to be followed by
experiments in browser performance (section \ref{sec:browser}) and
scaling behavior (section \ref{sec:scaling}). Finally, the last
section will present conclusions and future lines of research.

\section{State of the art}
\label{sec:soa} 

So called  {\em volunteer
computing}~\cite{sarmenta-sabotagetolerance,sarmenta-bayanihan,hpvc} takes
advantage of the creation of an infrastructure so that different people
can donate CPU cycles for a joint computing  effort.
The best known project is SETI@home\footnote{See
\url{http://setiathome.berkeley.edu/} for downloading the software  and some
reports.}, which, from the user's point of view is a screen-saver which has to be
downloaded and installed; when the user's CPU is not busy it performs
several signal analysis operations.
Some companies related to volunteer computing, such as Popular Power (and
others; they are referenced,
for example, in~\cite{Cappello}) did some experimentation with Java based
clients, but none has had commercial success.

There are mainly two problems in this kind of networks: first of all, it is
important not to abuse volunteers CPU resources; secondly, a sufficient
number of users is needed in order to be able to do the required
computation;
this can also be a problem on its own if there are too many users for the considered
setup. A third problem is that performance prediction is difficult
when neither the number of participants nor their individual node
performances are known in advance. 
In any case, we believe that the best way to obtain enough users is to make
it easy for them to participate, using technologies available in their
computers, as the browser is.
In fact, some  suggestions have been published (for example, the one of 
 Jim Culbert in his
weblog \cite{ajax:dc}, and in some mailing lists), but we are not aware of
any serious study about it.

The proposed approach could also be considered as {\em parasitic computing}
since, as stated in Section~\ref{sec:intro}, the only participation of the
user will be to load a web page; in fact, it could use these resources
without his acquiescence (and, in any case, it would be desirable to run
without causing much trouble).
The concept was introduced by Barab? in~\cite{Barabasi2001Parasitic}, and followed by others (for instance,
Kohring in 
\cite{kohring-2003-14})
In that work they proposed to use the Internet routers to compute a 
{\em checksum} by means of a set of specially crafted  packets to solve the
SAT problem. Anyway, although the concept is interesting, there seems not to be a continuation for this work.

The virtual machine embedded into the browser provides a way to easily
do that kind of  sneaky/parasitic computing, but JavaScript has the
problem to be an interpreted language and the efficiency of different
implementations varies wildly. Moreover, it is not optimized for
numerical computation but for object tree 
management (the so called DOM, document
object model) and strings.
Nevertheless its wide availability makes us think about considering it, at
least as a possibility.
It is also  important to remember that these resources can be used without
the user's participation (they only need to visit a web-page) opens a wide
set of alternative possibilities (and dangers, of course).

In this work an evolutive computation system will be presented. 
it has been developed in a Ruby on Rails based framework that takes
advantage of this feature; in this sense the approaches is new. We
will tackle the three main problems: abundance of clients (via the
system itself), possibility of client CPU abuse (also via de system
itself; the JS virtual machine runs within a sandbox inside the
browser), and performance prediction (which we will try to approach
via several experiments and benchmarks). 

\section{Resources and methodology}
\label{sec:met}

For the experiment we need several clients with JavaScript equipped browsers
and a server running  Ruby on Rails. RoR applications include their
own web server, WEBrick, but there are other options, such as
Mongrel\footnote{\url{http://mongrel.rubyforge.org/}} and {\sf
lighttpd}\footnote{\url{http://lighttpd.net/}} that are faster and will be
preferred for the experiments described below. 

The application follows MVC model; in this sense it is organized as a
model, a view and a controller.
We will concentrate on the first and the last of them.

The {\bf model} is a table in the database representing the population,
that could  be similar to this\footnote{In fact, it stores also some
information about the algorithm it belongs to: an identifier and the
state}:\begin{verbatim}
create table guy {
  cromosoma varchar(256),
  fitness float
};
\end{verbatim}
This table will store the population. 
We have preferred a traditional representation using a binary string; the
chromosome will be a list of 0s and 1s.
We are also using a scalar fitness, represented with a single {\sf
float} value\footnote{Which limits us, for the time
being, to single--objective optimization; but, in fact, there is no
constraint, since current database management systems can work with vectorial data types}. In any case, the data model is
related to the application we want to optimize and different chromosomic
representations and different fitness would need a different data model.

For the {\bf controller}, we will need controls that request new elements
from the population pool and to re-send them once evaluated.
This is also related to the labor division among clients and server.
We need to take into account that the evolutive algorithm needs to include
several actions: evaluation, genetic operators (mutation and crossing), and
merging of the population. 
The simplest way of task distribution is to evaluate fitness parameters (it
is usually the most time consuming operation) on the clients and to do the
other steps on the server, as shown in the algorithm; 
this scheme is usually called {\em farming}.  
Technically, the evaluation step should be included in the {\em view},
since it is interpreted in the client; in practice, it will be a JavaScript
program that will be included in the templates stored in the corresponding
directory of the RoR application.
Obviously, the fact that it will be executed on the client has security and
authentication consequences that have to be considered (including,
probably, fraud as shown in~\cite{sarmenta-sabotagetolerance}).
Since we are doing our experiments in a controlled way, they have not been
considered. We will only consider IP-based authentication (that is, in
this experiment we know in advance which IP addresses are going to
participate in it), and we will suppose that clients
will not send a higher fitness than the computed one (which would give false
results). Of course, there are other  methods to deal with this, such
as replicating evaluations in different clients (and comparing them),
or using some kind of client/server codification that would hinder or
avoid tampering with data. 
Controls will be needed to generate individuals and for the genetic
operators. They will be written in Ruby because they will be executed on
the server.
The whole system can be sketched as follows, from the client's point of
view:

\begin{enumerate}

\item Loading of the client code, which will be done along with
the web page, identified by an experiment unique URL: it will start
when the web page is loaded on the browser (by 
means of {\sf onLoad} browser's event) or at the user's request.

\item Request individuals to the server in order to evaluate them.  The
server sends a prefixed number of individuals  (a {\em package}).  If there
are not enough individuals to be evaluated, they will be generated on the
fly (by applying genetic operators).

\item The client evaluates individuals and send the result back to the
server. It will be evaluated in the server by the controller's method
  {\sf populationReady}. 
 Several formats can be used for the information interchange. Being AJAX
 the selected technology, it would seem natural to use XML, but we
 selected JSON ({\em JavaScript Object Notation}). 
 JSON\footnote{More information in \url{http://www.json.org/}} is an object
 serialization protocol that uses alphanumeric strings for data structures.
 It can be evaluated in JavaScript in order to convert it to an object and
 Ruby can also interpret and produce it from the database in a very
 straightforward manner.

\item The server uses tournament selection to generate new
individuals. In this method from a number  $n$ of individuals, the worst $p
< n$ are suppressed and substituted with the  offspring of the rest of
individuals; of course it can also be done selecting the best ones until
the number of needed individuals is reached (usually the same number as in
the original population). 

This tournament can be done in several steps by means of random
tournaments that will serve to detect the worst individuals and to 
eliminate them; then we will take the remaining ones in order to reproduce
them.  
The algorithm terminates when a number of individuals has been evaluated or
when a prefixed fitness level is reached.  In this way a percentage of the
new individuals will be generated by means of the available genetic
operators.  
\item The reply is sent to the client where a {\em callback} is generated
in order to return to the evaluation step with these new individuals.
\item The algorithm terminates when the client stops or when some condition
is reached; for example, the prefixed fitness is reached, or the selected number
of evaluations has been made.

\end{enumerate}

\begin{figure*}[htb]
\begin{center}
\includegraphics[scale=0.4]{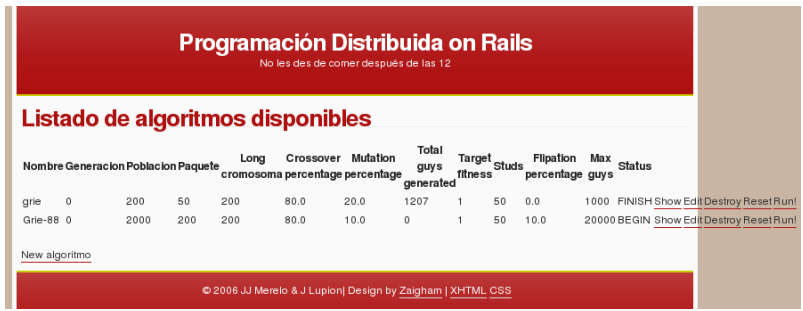}
\caption[Algorithm]{Screen capture of the  DCoR application running on the
browser. It shows different controls and algorithm parameters which will be
executed in the tests.\label{fig:algoritmos}}
\end{center}
\end{figure*}

\begin{figure*}[htb]
\begin{center}
\includegraphics[scale=0.5]{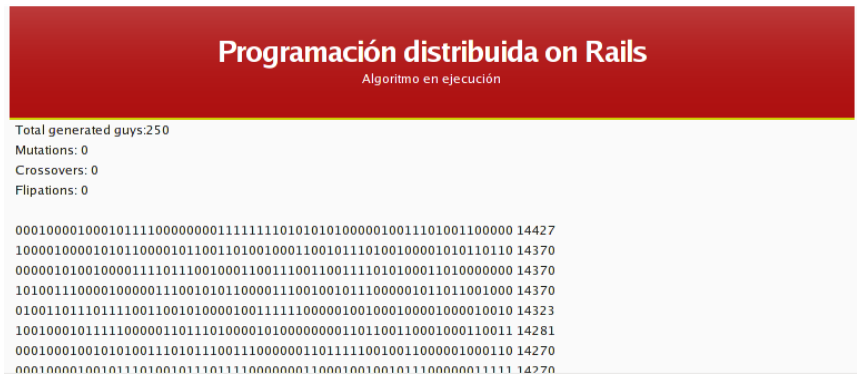}
\caption[Running]{The algorithm is executing. It shows the content of each
individual and the fitness value for the 64 bits binary knapsack
problem.\label{fig:ejecutando}} \end{center}
\end{figure*}

The parameters and the execution of the algorithm are configured by means
of the web page, as shown in Figure~\ref{fig:algoritmos}.
The algorithm can be executed by clicking on Run from its web page,
where it is also possible to modify the parameters or to restart it,
as is shown in the figure mentioned above.  In any
case, each algorithm has its own URL in a format like
\url{http://node:3000/algorithm/generation/<algorithm ID>}, that can be
used to run it from any browser. A screen capture of an algorithm running is shown 
 in Figure~\ref{fig:ejecutando}. 
The web page is dynamically refreshed when new server requests are
received.
The source code of the project is at RubyForge\footnote{
\url{http://rubyforge.org/projects/dconrails/}}, and it has been licensed
under the GPL; notice that this is an ongoing work and the software
state at each moment can or cannot correspond to the ideas expressed in this
paper.

%*******************************************************
\section{Browser performance}
\label{sec:browser}
%*******************************************************

Several experiments on different
browsers and with a limited number of computers
(\cite{maeb2007:dcor,dconrails-jp}, in Spanish),  have yielded
the result that different browsers have very different JavaScript
virtual machine performance, with Opera consistently outperforming the
rest, and Konqueror (the default KDE browser) coming last in
performance. With ad-hoc computation you can't choose the computer
the program is going to run eventually on, but it's always interesting
to have this data at hand when trying to predict the performance of a
particular problem, or estimate how much time a problem is going to
take based on statistics of browser usage in a particular web site. 

Performance also varies with the kind of problem (specially depending
on the kind of operations used to compute fitness, and the data types
--integer or floating point--), so, in this
occasion, a floating-point based problem has been chosen: the
10-variable Griewank \cite{griewank}
function, 
\begin{equation}
F(x) = \sum_{i=1}^n \frac{{x_i}^2}{4000}+\prod_{i=1}^n \cos{ \frac{x_i}{\sqrt{i}}} + 1
\end{equation}
$x_i$ is in the range -511,512. This function is characterized by a
high number of local minima, although it can be easily solved using any global
optimization procedure. We are not really interested in its
difficulty, but in the fact that it has got a size and a complexity
adequate to measure performance. In our experiments, we have chosen
$n=10$. The chromosome uses 20 bits to encode each floating point number, so that
each {\em gene} is decoded by computing $x_i =
(M-m)\frac{c_i}{1048575}+m$, where $M$ and $m$ are the range minimum
and maximum, and 1048575 the biggest number that can be coded with 20
bits and $c_i$ is the binary value of the gene. 
\begin{figure}[htb]
\begin{center}
\includegraphics[scale=0.4]{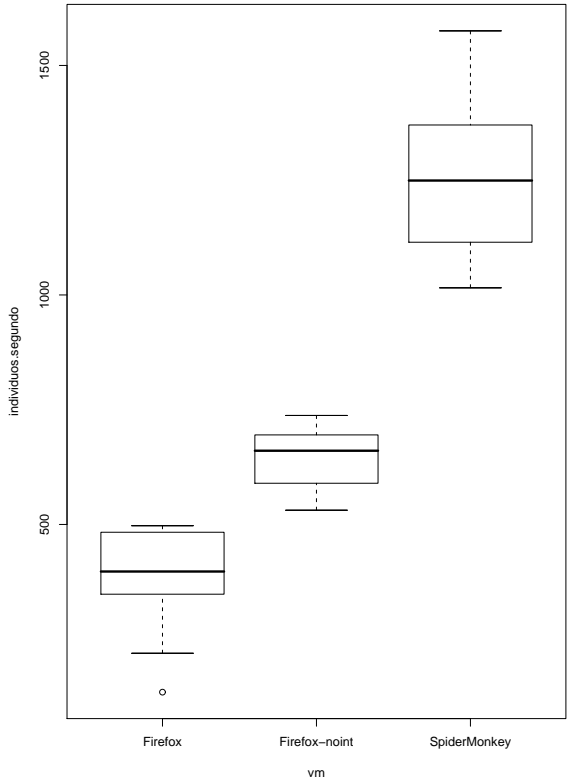}
\caption[Benchmark]{Boxplot of the number of evaluated chromosomes
per second, for three different JS virtual machines: the
SpiderMonkey stand-alone VM, Mozilla Firefox with script running time extended
to 30 secs, and Firefox with running time limited to 10 secs (default
setting).\label{fig:vms}} 
\end{center}
\end{figure}

The experiments using this setup has been carried out as follows:
first we have measured the individual evaluation rate in several 
configurations: a stand-alone Javascript interpreter ({\tt
JavaScript-C 1.5 2004-09-24} running on a Fedora Core 5 and a 
AMD Athlon(tm) 64 X2 Dual Core Processor 4200+), and Firefox\footnote{{\tt
Mozilla/5.0 (X11; U; Linux x86\_64; es-ES; rv:1.8.0.8) Gecko/20061108
Fedora/1.5.0.8-1.fc5 Firefox/1.5.0.8}} with two
different settings: the default setting, and the {\em no-int}, which
allows JavaScript programs to run without interruption for 30
seconds. The scripts were run several times (5 to 15) in the same machine
with the usual, and similar, workload. Results in number of
individuals evaluated per second are shown in figure \ref{fig:vms}.

These measures give us a baseline, without any intervention of the
evolutionary algorithm, of a few hundred to one thousand chromosomes
evaluated for fitness, per second, for this particular
problem. Incidentally, it also indicates that the browser architecture
and settings have a high impact on performance, which will have to be
taken into account when trying to predict performance for a particular
setting in advance\footnote{Preliminary results with the Opera
browser, not shown here, would be closer to the SpiderMonkey VM than
to Firefox in either configuration}.

The second experiment will try to measure the impact of packet size on
overall performance. Chromosomes are sent to the browser in packets of
$n$ individuals, which are then decoded, evaluated, and sent back to
the server (just the ID and fitness). This takes some time, and generates overhead in the shape
of database requests, data structure conversion, and the trip back and
forth itself (latency). Initially, bandwidth is not an issue, at least from the
client point of view, since these tests take place  on a local area
network (two computers connected to the same domestic ADSL router
through an Ethernet 100Mbit/s connection). 

\begin{figure}[tb]
\begin{center}
\includegraphics[scale=0.4]{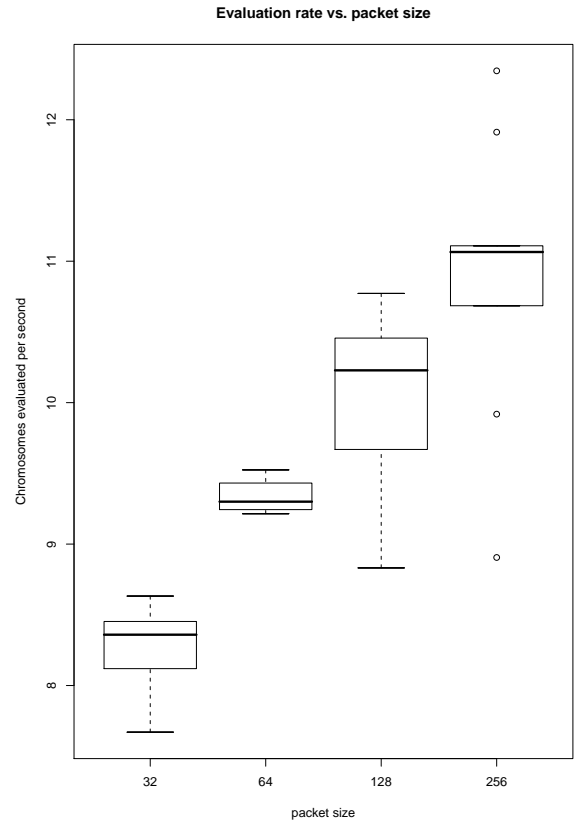}
\caption[Packet]{Boxplot of the number of chromosomes whose fitness
is evaluated in 1 second, depending on packet size in a client/server
setup. Clearly, performance increases with packet size, which implies
that latency and other overhead have a measurable impact on overall
performance.\label{fig:packets}} 
\end{center}
\end{figure}

Experimental setup is as follows: the server runs in the same computer as above, while the
client runs in a Sony VAIO VGN-S4XP with an Intel Pentium M (2 GHz)
running Firefox on Ubuntu 6.06, upgraded to the latest version (Jan
2007). An evolutionary algorithm that used 80\% crossover and 20\%
mutation rate, a population of 512 with an elite (extracted for
reproduction) of 256, and packet sizes of 32, 64, 128 and 256 was run
several times. The
number of evaluations was set to 5000, but, since the packet size is
not a whole multiple of that amount, the simulation usually ended with
a few more individuals evaluated. These were taken into account when
computing the chromosome evaluation rate, shown in figure
\ref{fig:packets}.
This figure, which can be fitted by the lineal model $n(s) = 8.36 \pm
0.19 * 0.011 \pm 0.001 s$ with 99.9\% confidence, shows that a new
chromosome can be evaluated for every 100 that are added to the
packet, and is obviously related to the number of petitions. A
doubling of packet size slashes by half the number of request and
responses from the server, decreasing also the number of database
queries. 

On the other hand, this model predicts that to achieve a
performance similar to the figures shown above (500-1500
chromosomes/second) the packet size would have to be an unreasonable
90000; 100 chroms/second could be achieved with a packet sized around
8000. This hints at a way of squeezing more
performance out of this setup (for instance, using a network with low
latency, or increasing the speed of the database), but also points to
a problem: bigger packet sizes means the client will be busy for more
time, during which the client could just turn the computer off or
wander away to another web page. In any case, it indicates that packet
sizing will have to be considered carefully in browser-based
distributed evolutionary computation, and also that the promises of
massively parallel performance will only be achieved when more than 20
computers are used to solve the problem. That is why it is interesting
to see how the server, which is the bottleneck in scaling, behaves
when the number of concurrent clients increases. This is what we are
going to do next. 

%*******************************************************
\section{Scaling behavior}
\label{sec:scaling}
%*******************************************************

In order to perform this experiment, an assortment of different
computers, with speeds ranging from 750MHz to 2.8 GHz, were added, one
by one. Packet size was set to 100 and equal to population size; the
rest of the parameters are irrelevant. Connections also varied from a
local connection in a 2-processor computer, to Fast Ethernet to
WiFi. Experiment did not start at the same time in all computers, but
more or less sequentially (actually, some people had to physically set
the browser URL). Besides, very few of
them were fully dedicated to the task; the URL was loaded while other
people were working on the computer. That is why  an improvement in averages should not really be
expected, but we should expect, at least, an improvement in the best
case, when all computers are started in a short period, and there is
no net congestion or CPU overload in any of them. In every case, experiment was
repeated several times. 

\begin{figure}[t]
\begin{center}
\includegraphics[scale=0.4]{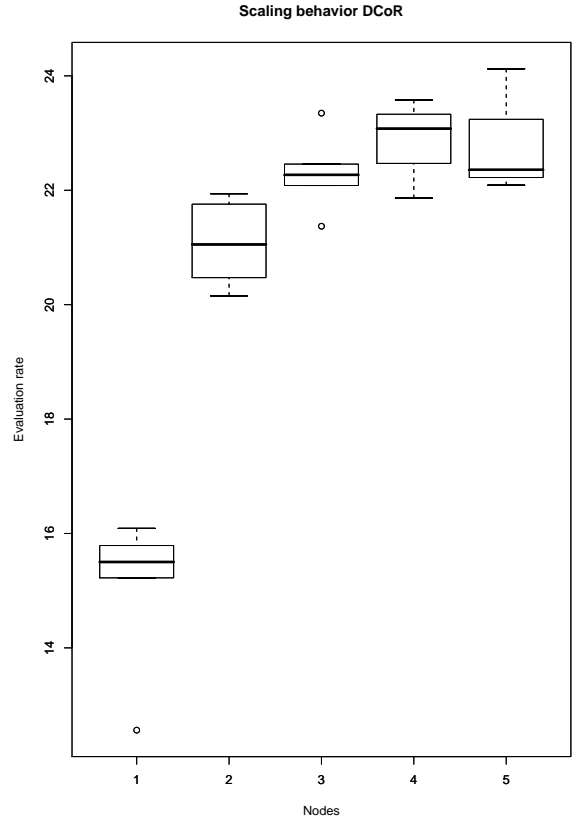}
\caption[Packet]{Boxplot of evaluation rate for a single server, and
several clients, ranging from 1 to 5.\label{fig:scaling}} 
\end{center}
\end{figure}

That is what can be
observed in figure \ref{fig:scaling}, which shows the boxplot of the
evaluation rate (total number of chromosomes evaluated divided by time
in seconds, as measured from the information stored in the server log)
vs. number of nodes. 
Scaling is dramatic from in the first steps, but it lowers down when 4
or 5 computers are added. Best-case result always improves, but not
dramatically, and average improvement slows down to a halt. This is
due to a number of factors, not the least being that the last computer
added was one of the slowest, but also to the fact that the server is
not running in full production mode, and is spending some time logging
information. There is also an architecture problem with the RoR
server: while all requests run in its own thread, the request/response
loop is sequential, so that just a single response can be served to
the client simultaneously. 

\begin{figure}[htb]
\begin{center}
\includegraphics[scale=0.8]{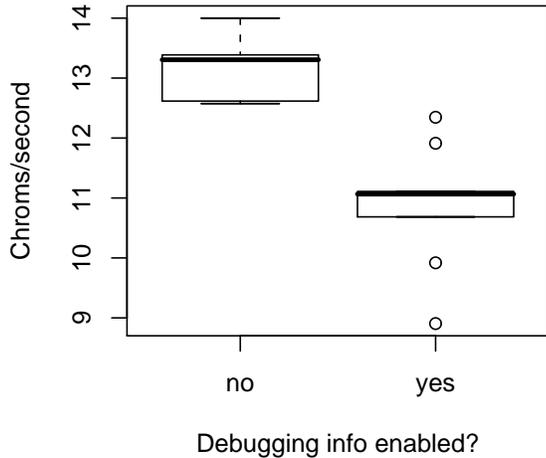}
\caption[Packet]{Boxplot of evaluation rate for a single server and
client, in production mode and with debugging messages to log-file
turned off (left), and in development mode with debugging messages on (right).\label{fig:nodebug}} 
\end{center}
\end{figure}

However, what is really important is the fact that we can
obtain a best-case improvement, as we are showing here. If debugging
is turned off (which we haven't done here for the full experiment
since the information is needed to check that the algorithm is running
correctly), some improvement can be obtained, as is shown in figure \ref{fig:nodebug}

The experiment shown in figure \ref{fig:nodebug} was done using the
same algorithm configuration as in \ref{fig:packets}, and an
improvement of around 20\% on average is obtained. This has an added
advantage: logging messages are sequential, which means that threads
handling a client must wait until other threads finish writing in that
single file, provoking interlocking problems, which might also account
for the poor average performance improvement observed above. We will
try to fix this in the next version of the DCoR package. 

\section{Conclusions and future work}

The main purpose of this paper has been to introduce the DCoR
(Distributed Computation on Rails) framework, make some measures to
find out what kind of performance we can expect from it, and run
scaling experiments on a simple configuration to highlight scaling
problems with it. We conclude that, barring major optimization and
tweaking of server performance, and using in each case the best
browser/client combination available, a good amount of clients is
needed to equal the performance of a stand-alone machine running the
same algorithm. But the problem is that, in the current setup, using a
multi-threaded but single-process server, that amount of scaling
cannot be achieved, with performance peaking when a few clients are
added. 

This leaves several possible paths for improvement: making DCoR fully
reentrant, so that multiple copies can easily run at the same time in
a server, and using a configuration of server clusters with a reverse
proxy (which is not trivial, but not too difficult either), or
changing the DCoR model so that more computation is moved to the
clients, leaving the server as just a hub for information interchange
among clients; that information interchange will have to be reduced to
the minimum, and, if possible, a single chromosome per
generation. That will make this model closer to the island model,
being every browser running the experiment a more or less independent
island, with just the migration policies regulated by the server. That
way, the server bottleneck is almost eliminated. In the near future,
we will try to pursue research along these two lines.

\section*{Acknowledgments}

This paper has been funded in part by the Spanish MICYT project  {\em NADEWeb:
Nuevos Algoritmos Distribuidos Evolutivos en la web}, code
{TIC2003-09481-C04}. We also acknowledge the support of the Spanish
Rails ({\sf ror-es}) mailing list.

%---------------------------------------------------------------------
\bibliographystyle{abbrv}
\bibliography{ror-js,GA-general,geneura,jini}

\end{document}